\documentclass[%
 reprint, 
 showkeys,
 amsmath,amssymb,
 aps, 
pra,
floatfix
]{revtex4-2}

\usepackage{graphicx}
\usepackage{dcolumn}
\usepackage{bm}
\usepackage{amsmath}
\usepackage{hyperref}
\usepackage[mathlines]{lineno}
\usepackage{color}
\usepackage{multirow}
\usepackage{booktabs}
\usepackage{physics}

\begin{document}

\title{A general approximator for strong-field ionization rates}

\author{Manoram Agarwal}
 \email{ manoram.agarwal@mpq.mpg.de}
\affiliation{%
 Max-Planck-Institut f\"ur Quantenoptik\\
Hans-Kopfermann-Str.\ 1\\
Garching 85748, Germany
}%
\author{Armin Scrinzi}
\email{Armin.Scrinzi@lmu.de}
\affiliation{Ludwig-Maximilians-Universit\"at M\"unchen\\
Theresienstrasse 37\\
80333 Munich, Germany
}
\author{Vladislav S.\ Yakovlev}
\email{vladislav.yakovlev@physik.uni-muenchen.de}
\affiliation{%
Ludwig-Maximilians-Universit\"at M\"unchen\\
Am Coulombwall~1\\
Garching 85748, Germany
}%
\affiliation{%
 Max-Planck-Institut f\"ur Quantenoptik\\
Hans-Kopfermann-Str.\ 1\\
Garching 85748, Germany
}%

\date{\today}

\begin{abstract}
We address the long-standing problem of determining accurate, time-resolved ionization rates for atoms in strong laser fields, a quantity that is fundamental to attosecond science. We show that it is possible to retrieve sub-optical-cycle dynamics of strong-field ionization from ionization probabilities obtained for a set of few-cycle laser pulses that covers a sufficiently broad parameter space. To this end, we introduce the General Approximator for Strong-Field Ionization Rates (GASFIR), a retrieval tool that uses a few adjustable parameters to accurately reconstruct \emph{ab initio} data. By imposing only essential physical constraints, our model provides a versatile framework for time-domain investigations of strong-field ionization and the role of ionization dynamics in attosecond metrology and lightwave electronics.
\end{abstract}

\maketitle

\section{Introduction}
The concept of the ionization rate is central to attosecond physics, serving as a foundation for analyzing how atoms and molecules respond to intense laser pulses. Accurate knowledge of ionization rates is essential for designing and interpreting experiments, as well as for controlling light-matter interactions at the petahertz scale. For example, recent advances in attosecond metrology enable time-domain measurements of optical fields using multiphoton ionization of atoms or photoinjection of carriers in a solid as a subfemtosecond gate~\cite{Sederberg2020, park2018direct, xiao_situ_2025}. The interpretation of such measurements heavily relies on theoretical ionization rates~\cite{NPS_theory, cho2021reconstruction}. Despite their important role, a rigorous description of ionization rates remains fundamentally challenging. Attosecond spectroscopy provides only indirect evidence of sub-optical-cycle ionization dynamics~\cite{goulielmakis_real-time_2010, nisoli_attosecond_2017, sie_time-resolved_2019}, and the very notion of ionization rate has been the subject of ongoing debate within the field as exemplified by ``attoclock'' measurements and their theoretical descriptions~\cite{pfeiffer_attoclock_2012, pfeiffer_recent_2013, sainadh_attosecond_2019}.

From a theoretical standpoint, there are two primary strategies for obtaining ionization rates. The first one is to search for approximations that result in analytical expressions for ionization rates, as is the case with the strong-field approximation \cite{bunkin_excitation_1964, keldysh1965ionization} and its refinements that account for the long-range Coulomb interaction \cite{faisal_multiple_1973, perelomov_ionization_1967, ammosov1986tunnel, REISS19921, lewenstein_theory_1994}. This approach has been very successful in understanding the physics of sub-optical-cycle non-adiabatic (henceforth, diabatic) ionization dynamics~\cite{yudin_nonadiabatic_2001, tagliamonti_nonadiabatic_2016, Weber_PRA_2025}, but by not being able to match the accuracy of \emph{ab initio} calculations, such approximate analytical theories may fail to incorporate important effects.

The second approach is purely numerical and avoids approximations that render analytical theories inaccurate. In the quasistatic limit, significant progress was made using complex scaling~\cite{scrinzi_ionization_1999,scrinzi2000ionization}. In the diabatic regime, the time-dependent Schrödinger equation (TDSE) is solved numerically for a given laser pulse. If it were possible to calculate instantaneous ionization probabilities from snapshots of an electron wave function, it would be possible to calculate ionization rates from time-dependent ionization probabilities.
However, all such attempts face the same fundamental problem: In the presence of a strong electric field, which exerts a force on an electron comparable to the force holding it within the atom, the distinction between the bound and continuum parts of the electron's wave function becomes ambiguous. It is possible to project the wave function onto the eigenstates of the unperturbed atomic Hamiltonian, but these states poorly represent the strongly perturbed Hamiltonian. It is possible to define time-dependent states that adiabatically evolve from the initial eigenstates, but this can be done in various ways~\cite{Karamatskou_PRA_2013,Yakovlev_CP_2013,Vabek_PRA_2022}, none of which guarantees an accurate calculation of instantaneous ionization probabilities. After the interaction with a laser pulse, an electron's wave function is easily separated into bound and continuum parts. However, since it is impossible to turn off ionization in the middle of a laser pulse while preserving all other quantum dynamics, it is impossible to track back what part of the final free-electron wave function was already free at a certain moment within the pulse.

So, there is a dilemma: Numerical calculations predict accurate ionization probabilities but yield ambiguous ionization rates. Analytical approaches provide explicit formulas for ionization rates but fail to achieve quantitative accuracy.
To resolve this dilemma, we propose to retrieve ionization rates from accurate (calculated or measured) ionization probabilities.
This idea is not entirely new: there are well-established approximations in the quasistatic limit \cite{Tong_2005}, where it is also possible to define and calculate an ionization rate as a functional derivative of the ionization probability \cite{ivanov_instantaneous_2018}.
Here, we propose a universal formula with five adjustable parameters that is suitable for retrieving ionization rates even when the quasistatic approximation fails.

\section{Theoretical background}
The most essential property of any ionization rate $\Gamma(t)$ is to predict the ionization probability via
\begin{eqnarray}
\label{eq:relPGamma}
    P_\mathrm{ion} = 1-e^{-\int_{-\infty}^{\infty}\Gamma(t')\,dt'}.
\end{eqnarray}
which follows from the rate equation for the concentration of neutral atoms: $d N_\mathrm{at} / dt = - \Gamma(t) N_\mathrm{at}(t)$.
Consider a hypothetical case where, for a certain atom, one can find a functional that maps the electric field $\mathbf{E}(t)$ of any laser pulse to such a function $\Gamma(t)$ that \eqref{eq:relPGamma} is always satisfied.
If such a functional were unique, there would be no objections to regard $\Gamma(t)$ as the rate of ionization even if instantaneous ionization probabilities during a laser pulse are fundamentally inaccessible.
However, the requirement of uniqueness requires from the functional more than correct ionization probabilities.
Here is a trivial example: the rates $\Gamma(t)$ and $a \Gamma(a t)$ predict the same ionization probabilities for any nonzero constant $a$.
What other condition must be imposed on the ionization-rate functional, which is essentially just a mathematical expression for $\Gamma(t)$ that contains $E(t)$?
Since ionization rates are unambiguously defined in the strong-field approximation (SFA), we propose the requirement that this mathematical expression must reproduce the known rates in the SFA limit. To prove the principle, we consider here an even stricter limit, where the Coulomb interaction of the free electron with the ion, the polarization of the ground state, the AC Stark effect, and transitions to excited bound states are neglected.
In summary, we define an ionization rate as the $\mathbf{E}(t) \rightarrow \Gamma(t)$ functional that predicts correct ionization probabilities and has the correct SFA limit.

In the SFA, the expression for the ionization rate can be written in the following form~\cite{lewenstein_theory_1994, NPS_theory}:
\begin{equation}
    \label{eq:kernel_definition}
    \Gamma(t) = \int_{-\infty}^{\infty}dT\,K(t,T),
\end{equation}
where the kernel $K(t,T)$ depends on $\mathbf{E}(t)$ in the time interval between $t-T$ and $t+T$. The time $T$ thus accounts for the non-adiabaticity of ionization.

In the following, we use atomic units unless otherwise stated. For simplicity, we assume that the laser pulse is linearly polarized along the $z$ axis: $\mathbf{E}(t)=\mathbf{e}_z E(t)$.
In this case, the SFA predicts (see Eq.~\eqref{eq:K_SFA} below)
\begin{equation}
    \label{eq:main_prop}
    K(t,T) =  E_{+} E_{\text{--}} \exp\left\{ i T \left[2 I_{\mathrm{p}} +\sigma_A^2(t,T)\right]\right\} f(t,T),
\end{equation}
where $I_\mathrm{p}$ is the ionization potential and $E_\pm  = E(t \pm T)$. We define the vector potential $A(t)$ by $E(t)=-A'(t)$, and Eq.~\eqref{eq:main_prop} uses the following notation:
\begin{multline}
    \label{eq:Var_A}
    \sigma_A^2(t,T) = 
    \frac{1}{2 T} \int_{t-T}^{t+T} A^2(t') \, dt' -\\ \left( \frac{1}{2 T} \int_{t-T}^{t+T} A(t') \, dt' \right)^2,
\end{multline}
which can be interpreted as the variance of the vector potential over the interval $2T$. Note that adding a constant to $A(t)$ does not change $\sigma_A^2(t,T)$.
The function $f(t,T)$ on the right-hand side of Eq.~\eqref{eq:main_prop} depends on the laser field and will be defined below.

According to the above equations, $\Gamma(t)$ depends on the values of the laser field before and after $t$. This noncausality emerges from the ambiguity in partitioning a wave function into bound and free components in the presence of a strong, time-dependent electric field.

Below, we outline the derivation of Eq.~\eqref{eq:main_prop} and propose a model for $f(t,T)$ that extends the applicability of Eq.~\eqref{eq:main_prop} beyond the SFA.

\subsection{The ionization-rate functional~\label{subsection:derivation}}
In the length-gauge SFA  (within the dipole approximation), the following expression (see Eq. \eqref{eq:SFA_der}) can be derived~\cite{lewenstein_theory_1994,NPS_theory}:
\begin{multline}
\label{eq:SFA_rate}
      K_\mathrm{SFA}(t,T) = 2 E_\mathrm{+}E_\mathrm{-}  e^{ i T \left(2I_\mathrm{p} + \sigma_A^2(t,T)\right)} \times \\
        \int d^3p\, e^{ i  T (\mathbf{p}+\mathbf{e}_z \overline{A})^2}
      d_z^*\bigl(\mathbf{p} + \mathbf{e}_z A_\mathrm{+}\bigr) d_z\bigl(\mathbf{p} + \mathbf{e}_z A_\mathrm{-}\bigr).
\end{multline}
Here, $A_\pm  = A(t\pm T)$, $\overline{A} = \frac{1}{2 T} \int_{t-T}^{t+T} A(t') \, dt'$, and $\mathbf{d(\mathbf{p})}$ is the dipole matrix element for the transition from the atomic ground state to the state where the atom is ionized, and the photoelectron has a momentum $\mathbf{p}$. 
The variable substitution $\mathbf{p}\to \mathbf{p} - \mathbf{e}_z \overline{A}$ transforms Eq.~\eqref{eq:SFA_rate} into
\begin{multline}
    \label{eq:K_SFA}
    K_\mathrm{SFA}(t,T) = 2 E_\mathrm{+}E_\mathrm{-}  e^{ i T \left(2I_\mathrm{p} + \sigma_A^2(t,T)\right)}
    \int d^3p\, \Biggl\{e^{ i  T p^2} \times \\
        d_z^*\biggl(\mathbf{p}+\mathbf{e}_z\frac{\xi_{2} + \xi_{1}}{2} \biggr) d_z\biggl(\mathbf{p}+\mathbf{e}_z\frac{\xi_{2} - \xi_{1}}{2} \biggr)\Biggr\},
\end{multline}
where we have introduced $\xi_{1} = A_+-A_-$ and $\xi_{2}= A_+ +A_- - 2\overline{A}$.
Note that for small values of $|T|$, $\xi_{1} \approx -2T E(t)$ and $\xi_{2} \approx -T^2 E'(t)$.
A comparison of Eq.~\eqref{eq:K_SFA} with Eq.~\eqref{eq:main_prop} reveals $f(t,T)$ in the SFA limit.

Our goal now is to generalize Eq.~\eqref{eq:K_SFA} beyond the SFA, while also making it suitable for ionization-rate retrievals.
We accomplish this goal by generalizing special cases where the integral over momenta in Eq.~\eqref{eq:K_SFA} can be evaluated analytically.
The details of this analysis can be found in Appendices~\ref{app:limits} and \ref{app:Gaussian}.
Here is the best general approximator for strong-field ionization rates that we have been able to find so far:
\begin{multline}
    \label{eq:fin_expr}
    f(t,T) = \pi^{3/2}a_0\left(\frac{i}{T+ia_1}\right)^{3/2+a_4} \times \\
       \left[\frac{2i  }{T+ia_1}-\xi_\mathrm{1}^2-\left(\frac{i T\xi_2 }{T+ia_1}\right)^{2}\right] \times \\
  \exp\left\{ -\frac{a_1}{4}\left( a_2\xi_\mathrm{1}^2+\frac{a_3 T\xi_2^2}{T+ia_1}\right)\right\}.
\end{multline}
In addition to the normalizing prefactor $a_0$, this model has four other adjustable parameters, from $a_1$ to $a_4$, that are supposed to incorporate not only the information that Eq.~\eqref{eq:K_SFA} stores in the dipole matrix elements but also the information required to describe the effect of the Coulomb force on the photoelectron and that of the laser field on the ground state prior to ionization.
In the next section, we validate GASFIR by showing that Eqs.~\eqref{eq:kernel_definition}, \eqref{eq:main_prop}, and \eqref{eq:fin_expr} allow for accurate and reliable retrieval of ionization rates from ionization probabilities within and beyond the strong-field approximation.

\section{Validation}
Numerical results in this section were obtained for laser pulses with the following vector potential:
\begin{equation}
    \label{eq:A}
    A(t) = \frac{E_\mathrm{L}}{\omega_\mathrm{L}} \cos^8\left(\frac{\pi t}{\tau_\mathrm{L}}\right) \cos(\omega_\mathrm{L} t)
\end{equation}
for $\pi |t| / \tau_\mathrm{L} \leq \pi / 2$ and zero outside this interval. Here, $E_\mathrm{L}$ is the peak electric field, $\omega_\mathrm{L}$ is the central angular frequency of the pulse, and $\tau_\mathrm{L}$ is related to the Full Width at Half Maximum (FWHM) of intensity via $\tau_\mathrm{FWHM} = 2 \tau_\mathrm{L} \arccos\left(2^{-1/16}\right) / \pi$.
The central wavelength of the pulse is $\lambda_\mathrm{L} = 2 \pi c / \omega_\mathrm{L}$, where $c$ is the speed of light.
The peak intensity of the pulse is $I = E_\mathrm{L}^2 / (2 Z_0)$ with $Z_0 = 377\,\text{ohm}$ being the impedance of free space.

For our SFA calculations, we numerically integrated over momenta in Eq.~\eqref{eq:K_SFA} and over $T$ in Eq.~\eqref{eq:kernel_definition}. That is, we did not use the saddle-point method to approximate any of these integrals. For the numerical solution of the single-electron TDSE in the dipole approximation, we used the tRecX code~\cite{scrinzi2022trecx}. With tReX, we calculated ionization probabilities using the time-dependent surface flux (t-SURFF) method~\cite{Tao_2012}.

\subsection{Reconstruction of ionization probabilities}
\begin{figure}[!htb]
    \includegraphics[width=\linewidth]{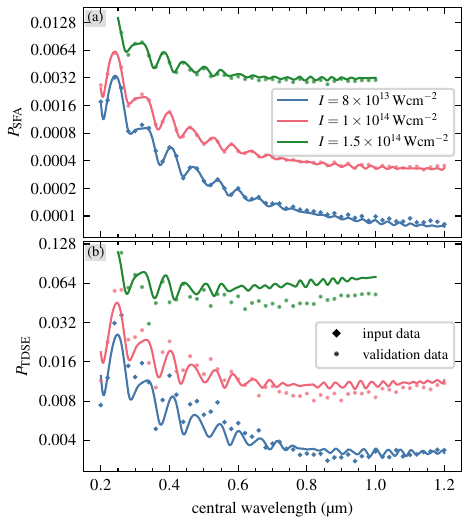}
    \caption{\label{fig:P_fit} GASFIR reconstruction of ionization probabilities.
    (a) Strong-field approximation. (b) The numerical solution of the TDSE.
    For both panels, the probability of ionizing a hydrogen atom with a two-cycle laser pulse was calculated for various combinations of the pulse's central wavelength and peak intensity. The retrieved probabilities (solid lines) are close to both the input data (diamonds) and the validation data (points).}
\end{figure}
We begin by demonstrating, in Fig.~\ref{fig:P_fit}, the accuracy with which GASFIR reconstructs ionization probabilities for two-cycle laser pulses ($\tau_\mathrm{FWHM} = 2 \times 2 \pi / \omega_\mathrm{L}$) in a broad parameter range typical for strong-field experiments.
For these tests, the input data consisted of a total number of 51 probabilities (shown by diamonds) calculated for a single value of the peak laser intensity: $8 \times 10^{13}\,\text{W}/\text{cm}^2$.
We used numerical optimization (further details are provided in Supplemental Materials \cite{SM} and in \cite{Edmond_GASFIR}) to find the GASFIR parameters that best reproduce this data, see the first two rows of Table~\ref{tab:fit_params}.
\begin{table}[!b]
\caption{%
Fit parameters obtained by fitting on the input data obtained from SFA and TDSE calculations}
\begin{ruledtabular}
\begin{tabular}{crrrrrrr}
& 
$a_0$ & $a_1$ & $a_2$ & $a_3$ & $a_4$  \\
\colrule
        H${}_\mathrm{SFA}$ & 3.38  & 3.50 &2.04 &0.28 &0.00  \\
        H${}_\mathrm{TDSE}$ & 6.51  & 3.50 &3.41 &3.19 &-0.50    \\
        H${}_\mathrm{QS}$ & 2.84  & 3.50 &3.88 & NA &-0.50  \\
        He${}_\mathrm{QS}$ & 0.02  & 1.02 &2.21 & NA &-2.91  \\
        Ne${}_\mathrm{QS}$ & 0.93  & 1.91 &4.17 & NA &-0.03   \\
\end{tabular}
\end{ruledtabular}
\label{tab:fit_params}
\end{table}
In the same plots, we compare another set of ionization probabilities (circles) against the GASFIR predictions.
For both SFA and \emph{ab initio} data, the GASFIR results are close to the training and validation data. The quality of agreement is worse for TDSE indicating that Eq.~\eqref{eq:fin_expr} only partially accounts for the role of the Coulomb force that the ion exerts on the photoelectron, as well as for the polarizability of the ground state (AC Stark shift). These effects influence channel closing~\cite{milosevic_wavelength_2008}, which we clearly see in both panels of Fig.~\ref{fig:P_fit} as the oscillations of the ionization probability with respect to $\lambda_\mathrm{L}$.
Furthermore, the maximal intensity in this test ($1.5 \times 10^{14}\,\text{W}/\text{cm}^2$, green curve) exceeds the barrier-suppression intensity, which also violates the assumptions of the SFA derivation. Nevertheless, GASFIR's predictions are qualitatively correct.

\subsection{The reliability of retrieved ionization rates}
\begin{figure}[!b]
    \includegraphics[width=\linewidth]{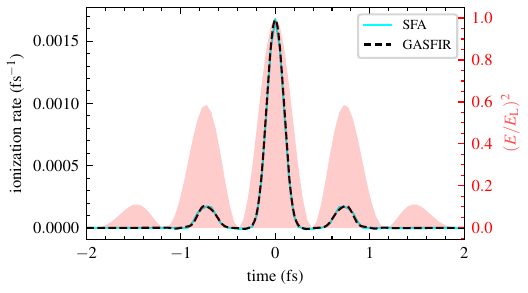}
    \caption{GASFIR correctly retrieves the SFA ionization rate (blue) from the SFA ionization probabilities.
    The SFA rate was calculated using Eq.~\eqref{eq:K_SFA} without any adjustable parameters for a hydrogen atom exposed to a 500-nm, single-cycle laser pulse with a peak intensity of $I=10^{14}$ W$\,$cm$^{-2}$ ($E_\mathrm{L}=2.75$ V$\,$\AA$^{-1}$).
    The retrieval process did not use the SFA rates.
    The shaded area represents the square of the electric field.}
    \label{fig:rates_compare_SFA}
\end{figure}

In the strong-field approximation, we can directly compare ionization rates retrieved by GASFIR with those calculated by numerically integrating Eqs.~\eqref{eq:K_SFA} and Eq.~\eqref{eq:kernel_definition}.
We observed excellent agreement, and Fig.~\ref{fig:rates_compare_SFA} shows a typical example for a single-cycle pulse ($\tau_\mathrm{FWHM} = \pi / \omega_\mathrm{L} = 1.67 \text{ fs}$).
We thus see that both requirements that we formulated in our definition of the ionization rate are satisfied: the accurate reconstruction of \emph{ab initio} ionization probabilities that we see in Fig.~\ref{fig:P_fit}(b) combined with the accurate reconstruction of the SFA rates. Nevertheless, let us see if there is additional evidence that $\Gamma(t)$ retrieved from \emph{ab initio} data indeed has properties that one would expect from an ionization rate.

For this purpose, we consider the following time-dependent quantity accessible in a numerical solution of the TDSE: the probability to observe an electron at a distance from the ion that is sufficiently large to regard the electron as free~\cite{bauer1999exact}. This quantity is not exactly the instantaneous ionization probability because for any choice of the cutoff distance $r_0$ that safely excludes the ground state, some parts of a free-electron wavepacket need a significant time to reach the distance, and no choice of $r_0$ can exclude all Rydberg states. For this reason, we cannot directly compare this probability, $P_{r\geq r_0}(t)$, with instantaneous probabilities obtained by integrating GASFIR rates. For a fair comparison, we combine GASFIR rates with a classical analysis of electron trajectories that accounts for the Coulomb force that the ion exerts on the photoelectron, see Appendix~\ref{sec:classical} and \cite{NPS_theory} for details.
A similar approach has been utilized to reconstruct ionization rates from TDSE predictions~\cite{teeny_ionization_2016}.

\begin{figure}[!t]
    \includegraphics[width=\linewidth]{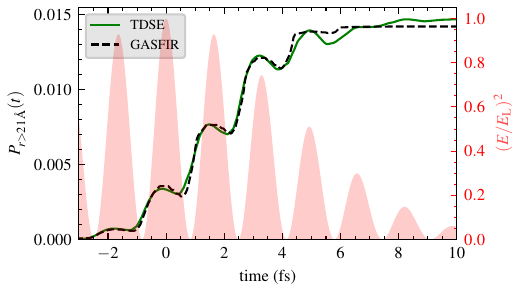}
    \caption{The probability of finding an electron at a distance $r \geq 21$~{\AA} from the ion as a representation of ionization dynamics. Here, we compare the probability obtained from the numerical solution of the TDSE to that evaluated by combining GASFIR's ionization rates with a classical analysis of electron trajectories. The laser parameters were: $\lambda_\mathrm{L}=1000$~nm, $I=10^{14}$~W$\,$cm$^{-2}$ ($E_\mathrm{L}=2.75$ V$\,\text{\AA}^{-1}$), $\tau_\mathrm{FWHM}=3 \times 2\pi/\omega_\mathrm{L} = 10\,\text{fs}$.}
    \label{fig:position_compare_TDSE}
\end{figure}

Figure~\ref{fig:position_compare_TDSE} compares the ionization dynamics in both models.  Despite the simplicity of the semi-classical model, the agreement is remarkably good.
In both curves, the steepest increase is delayed with respect to the peak of the laser pulse because of the time it takes a photoelectron to reach a distance of 21~{\AA}.

\subsection{Quasistatic limit~\label{sec:QS}}
We now consider the quasistatic limit of GASFIR. This not only enables direct comparison with well-known theories of tunneling ionization but also provides a reference for detailed investigations of diabatic dynamics in strong-field ionization of atoms.
To derive an analytical expression for the quasistatic ionization rate using GASFIR, we consider the case of a constant electric field ($E(t) = E = \text{const}$) and use the saddle-point method to integrate over $T$ in Eq.~\eqref{eq:kernel_definition}, the integrand being defined by Eqs.~\eqref{eq:main_prop} and \eqref{eq:fin_expr}. The result is
\begin{multline}
  \label{eq:QS_kernel}
  \Gamma_\mathrm{QS}(E) = 2 \pi^2 a_0 |E| \frac{1 + 2 E^2 \tilde{T}_\mathrm{s}^2 (a_1 + \tilde{T}_\mathrm{s})}
  {\left(a_1 + \tilde{T}_\mathrm{s}\right)^{a_4+\frac{5}{2}} \sqrt{a_1 a_2 + \tilde{T}_\mathrm{s}}} \times \\
 \exp\left\{-2 I_\mathrm{p} \tilde{T}_\mathrm{s} + E^2 \tilde{T}_\mathrm{s}^2 \left(a_1 a_2 + \tilde{T}_\mathrm{s} / 3\right)\right\}, 
\end{multline}
where
\begin{equation}
    \tilde{T}_\mathrm{s} = \sqrt{\frac{2 I_\mathrm{p}}{E^2} + (a_1 a_2)^2} - a_1 a_2. 
\end{equation}
is the imaginary part of the saddle point $T_\mathrm{s} = i \tilde{T}_\mathrm{s}$.
Parameter $a_3$ is absent in $\Gamma_\mathrm{QS}$---it is related to $\xi_2$, which is zero for a constant electric field.

For moderately strong fields with $(a_1 a_2 E)^2 \ll 2 I_\mathrm{p}$, Eq.~\eqref{eq:QS_kernel} can be simplified to an expression that is similar to well-established analytical results for tunneling ionization:
\begin{multline}
    \label{eq:QS_appx}
    \Gamma_\mathrm{QS}(E) \approx  2 \pi^2 a_0  \frac{|E|^{3+a_4} \left(2 \kappa^3 + (2 a_1 \kappa^2+1)|E|\right)}{(\kappa + a_1 |E|)^{\frac{5}{2}+a_4} \sqrt{\kappa}} \times\\
    \exp\left\{-\frac{2 \kappa^{3}}{3 |E|}- \kappa (a_1 a_2)^2 |E| + a_1 a_2 \kappa^2+\mathcal{O}(E^2)\right\} 
\end{multline}
with $\kappa=\sqrt{2I_\mathrm{p}}$.
The first term in the argument of the exponential function in Eq.~\eqref{eq:QS_appx} is characteristic of the tunneling regime \cite{keldysh1965ionization,ammosov1986tunnel}, and it does not contain any of the adjustable parameters. 
Notably, a term equivalent to $e^{- \kappa (a_1 a_2)^2 |E|}$ appears in the empirical tunneling rate proposed by Tong and Lin~\cite{Tong_2005}, who introduced it to remedy the overestimation of rates in the barrier-suppression regime, $|E|>\kappa^4 / 16$.

\begin{figure}[!t]
    \includegraphics[width=\linewidth]{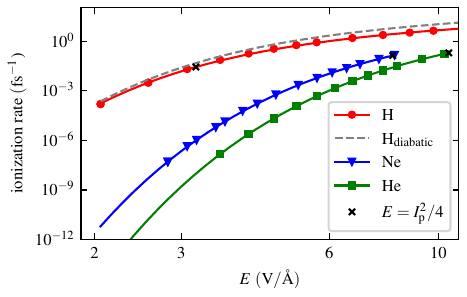}
    \caption{Tunneling (quasistatic) rates. The black crosses indicate the barrier-suppression fields. The other markers represent literature values calculated using the complex-scaling method for the ionization rate in a static electric field for atomic hydrogen (red circles), neon (blue triangles), and helium (green squares). The curves represent the quasistatic limit of GASFIR, Eq.~\eqref{eq:QS_kernel}. The solid curves were obtained by retrieving the model parameters from the data represented by the markers. The corresponding parameters are listed in the last three rows of Table 1. The gray dashed curve was calculated with the model parameters that best describe the ionization of atomic hydrogen by two-cycle laser pulses.
    }
    \label{fig:QS_rates}
\end{figure}
Figure~\ref{fig:QS_rates} compares previously published quasistatic (tunneling) rates for several atoms with ionization rates in the quasistatic limit of GASFIR.
The input data for hydrogen was calculated using the complex scaling method 
\cite{scrinzi2000ionization}.
The rates for helium and neon account for electron-electron interaction in the framework of the hybrid anti-symmetrized coupled channels (haCC) method~\cite{majety_photoionization_2015,majety_static_2015}.
For atomic hydrogen, the gray dashed curve represents the predictions of Eq.~\eqref{eq:QS_kernel} with the same model parameters as those we used to make Figs.~\ref{fig:P_fit}(b) and \ref{fig:position_compare_TDSE}.
To get the data for the solid curves in Fig.~\ref{fig:QS_rates}, we numerically optimized the parameters in Eq.~\eqref{eq:QS_kernel} to approximate the input data.
The model parameters are listed in Table~\ref{tab:fit_params}.
These results provide preliminary evidence that the GASFIR can describe the strong-field ionization of multielectron atoms.

\section{Conclusions}
For linearly polarized laser pulses, Eqs.~\eqref{eq:kernel_definition}, \eqref{eq:main_prop}, and \eqref{eq:fin_expr} provide a general framework (which we refer to as GASFIR) for retrieving multiphoton ionization rates from a set of ionization probabilities that can be measured in experiments or calculated \emph{ab initio}.
For atomic hydrogen, this model accurately reproduces correct ionization probabilities in a broad range of laser parameters, accounting for interference effects such as channel closing.
We have also obtained an analytical expression for the tunneling limit of our model for ionization rates, Eq.~\eqref{eq:QS_kernel}.
This expression very accurately approximates the quasistatic ionization rates of hydrogen, helium, and neon, which provides further evidence that GASFIR can account for the effect of the Coulomb interaction between the ion and the photoelectron on photoionization dynamics.
Although our demonstration of GASFIR is limited to atomic gases, the same approach is likely to work for molecules and other quantum systems. In particular, access to accurate photoinjection dynamics in semiconductors and dielectrics would significantly advance the field of lightwave (petahertz) electronics.

\appendix

\medskip
\textbf{Acknowledgments} \par 
The authors acknowledge fruitful discussions with F.~Krausz.
M.~A.\ acknowledges support from the Max Planck Society via the International Max Planck Research School of Advanced Photon Science (IMPRS-APS).

\medskip


%

\appendix
\renewcommand{\thesection}{\Alph{section}}
\section{SFA kernel\label{app:kernel_der}}
We start with the well-known equation for SFA's predicted continuum wavepacket~\cite{ivanov2005anatomy}  
\begin{multline} \label{eq:S-matrix}
   \lim_{t \to \infty} \ket{\Psi(t)} = -i \int d^3 p\, \ket{\bm{p}}\int_{-\infty}^{\infty} dt'\, e^{-\frac{i}{2}\int_{t'}^{\infty} [\bm{p}+\bm{A}(t')]^2 \, dt'}  \\e^{i I_\mathrm{p} t'}  
   \bra{\bm{p} + \bm{A}(t')} \bm{d} \cdot \bm{E}(t')\ket{\Psi_0},
\end{multline}
to get the ionization probability
\begin{multline}
\label{eq:SFA_der}
    P_\mathrm{ion} = \lim_{t \to \infty} \bra{\Psi(t)} \hat{1} \ket{\Psi(t)}
    \\=\int d^3p'\, \iint_{-\infty}^{\infty} dt_1\, dt_2\, \Bigl\{e^{i (t_2-t_1) I_\mathrm{p} +\frac{i}{2}\int_{t_1}^{t_2} dt'\, [\bm{p'} + \bm{e}_z A_z(t')]^2} \\E_z(t_2) d_z\bigl(\bm{p'} + \bm{e}_z A_z(t_2)\bigr) 
         E_z(t_1) d_z^* \bigl(\bm{p'} + \bm{e}_z A_z(t_1)\bigr)
 \Bigr\}.    
\end{multline}
We define $t=(t_2+t_1)/ 2$, $T=(t_2-t_1) / 2$, and express the phase-terms using the variance of the vector potential, $\sigma_\mathrm{A}^2(t,T)$. A simple comparison of the resulting integrand with Eqs.~\eqref{eq:relPGamma} and \eqref{eq:kernel_definition} gives Eq.~\eqref{eq:SFA_rate}.
\section{Limits~\label{app:limits}}
In addition to the numerical evidence that GASFIR successfully retrieves SFA rates, we can compare the analytical expression for $K_\mathrm{SFA}(t,T)$ and $K(t,T)$ in two limits: $T \to 0$ and $T \to \infty$. At $T=0$, Eq.~\eqref{eq:SFA_rate} reduces to
\begin{equation}
\label{eq:limT0}
    K_\mathrm{SFA}(t,T=0) = 2 E^2(t) \int d^3p\, \left|d_z\bigl(\bm{p} \bigr)\right|^2 \propto E^2(t).
\end{equation}
Equation~\eqref{eq:main_prop} with $f(t,T)$ from Eq.~\eqref{eq:fin_expr} yield the same result:
\begin{eqnarray}
    K(t,0) = \frac{2a_0}{a_1^{5/2+a_4}}E^2(t) \propto E^2(t).
\end{eqnarray}
In the limit of large $T$, the stationary-phase method allows us to integrate over $\mathbf{p}$ in Eq.~\eqref{eq:K_SFA}:
\begin{multline}
    \label{eq:limTinf}
     K_\mathrm{SFA}(t,T) \xrightarrow{T \to \infty} 2 E_{+}E_{-} e^{i T \left[2 I_{\mathrm{p}} + \sigma_A^2(t)\right]} \times \\
      \left(\frac{i\pi}{T}\right)^{3/2}
      d_z^*\biggl(\mathbf{e}_z \frac{\xi_{2} + \xi_{1}}{2} \biggr)
      d_z\biggl(\mathbf{e}_z \frac{\xi_{2} - \xi_{1}}{2} \biggr). 
\end{multline}
For GASFIR, we set $a_4=0$ because the SFA requires it. We also approximate $T + i a_1 \approx T$ for large values of $|T|$. This yields
\begin{multline}
     K(t,T) \xrightarrow{T \to \infty} a_0 \left(\frac{i}{T}\right)^{3/2}E_{+}E_{-} e^{ i T \left[2 I_{\mathrm{p}} +\sigma_A^2(t)\right]} \times \\
     e^{-\frac{a_1}{4}\left(a_2\xi_1^2+a_3\xi_2^2\right)} \left[\xi_2^2- \xi_1^2\right]. 
\end{multline}
Consequently, GASFIR reproduces $K_\mathrm{SFA}(t,T)$ in the limit of large $T$ if $\exp\left\{-\frac{a_1}{4}\left(a_2\xi_1^2+a_3\xi_2^2\right)\right\} \left[\xi_2^2- \xi_1^2\right]$ can approximate
$d_z^*\biggl(\mathbf{e}_z \frac{\xi_{2} + \xi_{1}}{2} \biggr) d_z\biggl(\mathbf{e}_z \frac{\xi_{2} - \xi_{1}}{2} \biggr)$.

\section{Analytical approximation of the integral over momenta~\label{app:Gaussian}}
The dipole matrix element describing transitions from the ground state of a hydrogen atom to a plane wave with momentum $\mathbf{p}$ is known to be
\begin{eqnarray}
    \label{eq:dipole_hydrogen}
    \mathbf{d}_\mathrm{1s}(\mathbf{p})=2^{7/2} (2 I_\mathrm{p})^{5/4} \frac{\mathbf{p}}{(p^2 + 2 I_\mathrm{p})^3}.
\end{eqnarray}
With this matrix element, it is impossible to analytically evaluate the integral over $\mathbf{p}$ in expression for $K_\mathrm{SFA}(t,T)$, such as Eq.~\eqref{eq:K_SFA}.
However, as long as the main contribution to the integral comes from small momenta (which is the case when ionization creates a wave packet with a small initial average velocity), we the following approximation for the bound-continuum matrix elements allows for analytical integration:
\begin{eqnarray}
    \mathbf{d}(\mathbf{p})=a_0 e^{-\frac{\tau p^2}{2}} \bm{p}.
\end{eqnarray}
In this case, we obtain
\begin{multline}
    \label{eq:SFA_1s}
    f_\mathrm{Gauss}(t,T)=\frac{a_0^2 \pi^{3/2}}{4}\left(\frac{i}{T+i\tau}\right)^{3/2}e^{-\frac{\tau}{4}\left(\xi_1^2+\frac{T\xi_2^2}{T+i\tau}\right)} \\
    \times \left[\frac{2i}{T+i\tau}-\xi_1^2-\left(\frac{i}{T+i\tau}\right)^{2}T^2 \xi_2^2\right].
\end{multline}
Equation~\eqref{eq:fin_expr} is a generalization of Eq.~\eqref{eq:SFA_1s}.



\section{Classical Newtonian propagation after ionization}
\label{sec:classical}
Let us consider the motion of a classical electron in the electric field of the light pulse, $\mathbf{E}(t)$, and the field of the singly-charged ion placed at $\mathbf{r}=0$. 
The force acting on an electron at position $\mathbf{r}$ is equal to
\begin{equation} \label{eq:force}
    \mathbf{F}(\mathbf{r}) = -\frac{\mathbf{r}}{|\mathbf{r}|^3} - \mathbf{E}
\end{equation}
in atomic units. By solving Newton's equation of motion, $\ddot{\mathbf{r}} = \mathbf{F}\bigl(\mathbf{r}(t)\bigr)$, we can find the position, $\mathbf{r}(t, t_0)$, of a free electron at time $t$ if its classical motion begins at time $t_0$. 
With that,
\begin{eqnarray}
    P_{r>r_0}(t)\approx \int_{-\infty}^t dt_0 \Gamma(t_0) \Theta(|\mathbf{r}(t,t_0)|-r_0).
\end{eqnarray}
Here, $\Theta$ denotes the Heaviside step function. We chose $\mathbf{r}(t_0, t_0)$ to be the point where the classically forbidden region ends along the line that passes through $\mathbf{r} = 0$ in the direction of the electric field, $\mathbf{E}(t_0)$.
The expression for this initial condition in atomic units is:
\begin{equation}
\mathbf{r}(t_0, t_0) = -\frac{\mathbf{E}(t_0)}{|\mathbf{E}(t_0)|}\frac{|I_\mathrm{p}|+\sqrt{I_\mathrm{p}^2-4|\mathbf{E}(t_0)|}}{2|\mathbf{E}(t_0)|}.
\end{equation}

To keep our classical model as simple as possible, we did not average over an ensemble of electron trajectories with different initial velocities.
Instead, we adjusted the initial velocity, which we took to be along the orthogonal axis.
The results shown in Fig.~\ref{fig:position_compare_TDSE} were obtained with $\mathbf{v}(t_0) = 0.5 \mathbf{e}_y$. We additionally provide a figure in Supplemental Materials that demonstrates how varying the choice of initial velocity by as much as 20\% does not affect the general conclusion that GASFIR ionization rates are quite meaningful. 

\end{document}